# Experimental detection of active defects in few layers $MoS_2$ through random telegraphic signals analysis observed in its FET characteristics

**Nan Fang**[*1], **Kosuke Nagashio**[1,2] **and Akira Toriumi**[1]
[1]Department of Materials Engineering, The University of Tokyo, Tokyo 113-8656, Japan
[2]PRESTO, Japan Science and Technology Agency (JST), Tokyo 113-8656, Japan
[*]E-mail: nan@adam.t.u-tokyo.ac.jp

**Abstract** Transition-metal dichalcogenides (TMDs), such as molybdenum disulfide ($MoS_2$), are expected to be promising for next generation device applications. The existence of sulfur vacancies formed in $MoS_2$, however, will potentially make devices unstable and problematic. Random telegraphic signals (RTSs) have often been studied in small area Si metal-oxide-semiconductor field-effect transistors (MOSFETs) to identify the carrier capture and emission processes at defects. In this paper, we have systemically analyzed RTSs observed in atomically thin layer $MoS_2$ FETs. Several types of RTSs have been analyzed. One is the simple on/off type of telegraphic signals, the second is multilevel telegraphic signals with a superposition of the simple signals, and the third is multilevel telegraphic signals that are correlated with each other. The last one is discussed from the viewpoint of the defect–defect interaction in $MoS_2$ FETs with a weak screening in atomically confined two-dimensional electron-gas systems. Furthermore, the position of defects causing RTSs has also been investigated by preparing $MoS_2$ FETs with multi-probes. The electron beam was locally irradiated to intentionally generate defects in the $MoS_2$ channel. It is clearly demonstrated that the $MoS_2$ channel is one of the RTS origins. RTS analysis enables us to analyze the defect dynamics of TMD devices.
**KEYWORDS:** $MoS_2$, random telegraphic signals, FET, defect-defect correlation, defect detection.

## 1. Introduction

Transition metal dichalcogenides (TMDs) have attracted much attention both in condensed matter physics [1,2] and in ultimately scaled device research [3,4]. It is surprising that 0.65 nm-thick monolayer $MoS_2$ can work well as a field-effect-transistor channel because conventional ultrathin Si metal-oxide-semiconductor field-effect transistor (MOSFET) performance is severely degraded by reducing its channel thickness [5,6]. Although the carrier mobility of $MoS_2$ is not as high as that of graphene, the off-current is definitely suppressed thanks to a sufficient energy band gap, which is very beneficial for device applications such as highly sensitive sensors and low-power FETs.

Defects such as sulfur vacancy can seriously affect carrier transport in terms of contact resistance [7], mobility and subthreshold swing measured by electrical measurement [2]. Moreover, defects also affect the noise behavior of $MoS_2$ devices. One of the studied noises in $MoS_2$ devices is $1/f$ noise, which is considered to be one of the severe problems in radio frequency applications [8-15]. The trap state density and trapping time of active defects can be extracted by applying number fluctuation theory for observed $1/f$ noise from $MoS_2$ devices [14,15]. Alternatively, the noise occasionally observed in conventional semiconductor devices is random telegraphic signals (RTSs) [16-19]. This refers to discrete current levels in a certain time domain. RTSs are often observed in sub-μm Si-MOSFETs and usually correspond to carrier capture and emission processes at defects. Although most fabricated TMD-FETs are not as small as small Si MOSFETs, their intrinsically thin body causes RTSs to be occasionally observed owing to a less effective screening effect in the two-dimensional system [20].

To the best of our best knowledge, the research of RTSs in TMDs is very limited; in particular, the position of noise origin, or the contact or/and the channel [10,11], is still under study. One of the observations is of RTSs in $MoTe_2$ FETs measured ambiently, where RTSs are considered to result from gas absorption/desorption at the $MoTe_2$ surface [21]. However, we have reported that RTSs can also be observed in $MoS_2$ FETs in vacuum, which suggests that the origin of RTSs might be the interaction of carriers with defects in a $MoS_2$ channel and/or some other parts [22]. If RTSs result from these defects, it may be possible to determine the defect position of RTSs using multi-probe measurements.

In this work, we systematically investigate the RTSs of back-gated thin (1–4 layers) $MoS_2$ FETs in vacuum. Many types of RTSs observed in this study are classified based on simple simulations using random numbers. Moreover, we demonstrate that intentional electron beam (EB) irradiation of the channel results in the clear observation of RTSs, which suggests that defects are introduced in $MoS_2$. Finally, to clarify the defect position of RTSs, multi-probe measurement was conducted. The results indicate that defects in the $MoS_2$ channel are among the origins of RTSs.

## 2. Results and discussion

In this study, $MoS_2$ was transferred on a 90-nm $SiO_2$/Si substrate by mechanically exfoliating natural bulk $MoS_2$ flake. Thin (1–4 layer) $MoS_2$ was selected as the channel material. Source and drain patterns were drawn by EB lithography, and metal (50-nm Ni / 50-nm Au) was thermally evaporated to form the contact. A Ti/Au electrode was used only for multi-probe devices. Figure 1a shows an optical image of a multilayer $MoS_2$ FET with multi-probes. Thin $MoS_2$ is clearly observed on the substrate. The layer number was confirmed by Raman spectroscopy [23], atomic force microscopy (AFM), photoluminescence (PL) [24] and photoconductivity measurements (details provided in supplementary figure S1).

More than 50 devices ranging from mono- to tetra-layer were measured to study RTS behavior. All *I*–*V* measurements were conducted by B1500 in vacuum at $1\times10^{-3}$ Pa to remove any absorption of moisture. First, typical *I*–*V* characteristics of a bilayer $MoS_2$ FET are shown in figure 1b. The results are shown as a parameter of the temperature from 20 to 300 K. The field-effect mobility based on four-probe measurement at room temperature for mono- and tetra-layer $MoS_2$ FETs are 14 and 26 $cm^2V^{-1}s^{-1}$, respectively. The estimated mobility is reasonable for back-gated $MoS_2$ FETs. Based on the temperature dependence of the subthreshold swing, the average interface state density was estimated to be approximately $4.5\times10^{12}$ $cm^{-2}eV^{-1}$ for monolayer devices (details provided in supplementary figure S2). Conductance fluctuations are clearly observed in the subthreshold region at lower temperatures. This conductance fluctuation was reproducible in each sample in repeated voltage sweeps. Here, we consider the percolation model, as shown in figure 1c, because conductance fluctuation is observed in the subthreshold region. From the similarity to graphene on $SiO_2$ [25], it can be expected that there is an electrostatic potential fluctuation in the channel due to charged impurities at the $SiO_2$ surface as well as in the $MoS_2$. The green squares are conductive paths, whereas white squares are resistive paths. Previously, defects such as sulfur vacancy in $MoS_2$ have been experimentally observed by scanning tunneling microscopy [26] and capacitance measurement [27]. The stability of sulfur vacancy was also simulated by the thermodynamic calculation [28]. It is suggested that a specific defect in the $MoS_2$ channel and/or at the $SiO_2$ substrate surface may be responsible for each conductance fluctuation. We assume that defects are randomly distributed throughout the channel and that one of them locates at the critical point of percolation path, as schematically shown by an arrow in figure 1c. In this case, the interaction of carriers with the defect located at the critical point of percolation path could result in the conductance fluctuations [29], because this critical defect determines the total resistance.

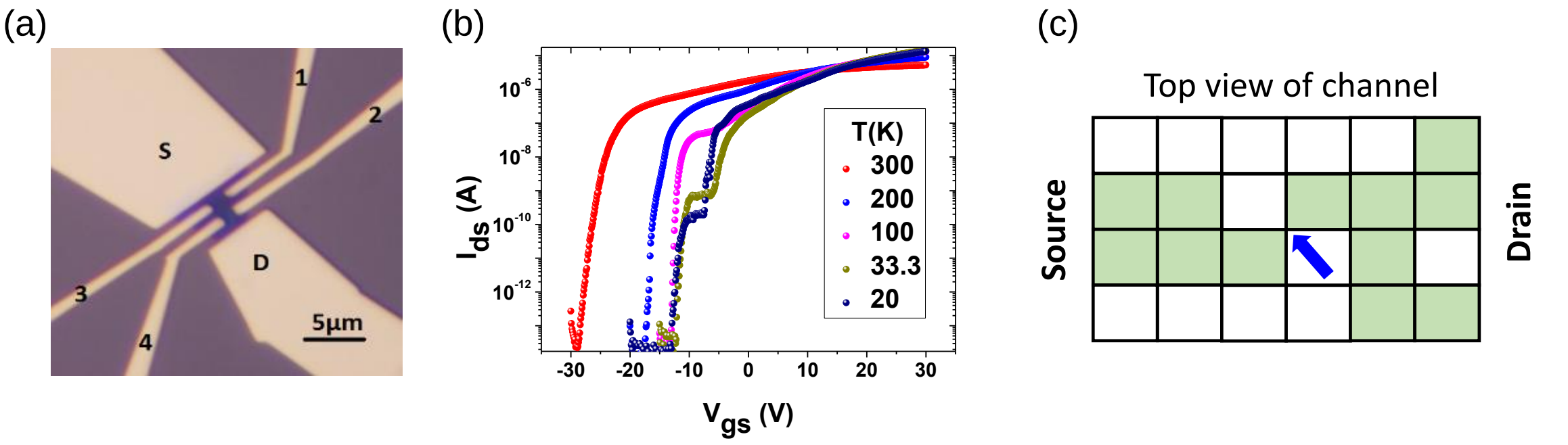

**Figure 1.** (a) Optical image of multi-probe 3–4 layers of $MoS_2$ on 90-nm $SiO_2$/Si substrate. (b) Subthreshold transport characteristics in bilayer $MoS_2$ FET ($V_{ds}$ = 0.1 V). (c) Schematic drawing of the channel (top view), showing the percolation condition. Green squares are conductive paths, whereas white squares are resistive paths. This conductivity inhomogeneity comes from surface potential inhomogeneity. The blue arrow indicates the critical path because this point connects the source and drain and determines the total resistance. If there is a defect around this critical point that can communicate with a carrier in $MoS_2$ through the carrier capture and emission process, conductance fluctuation can be observed.

Figure 2a shows simple two-level RTSs observed in bilayer $MoS_2$ measured at 20 K. RTSs are often observed by measuring the drain current ($I_{ds}$) as a function of time at a fixed gate voltage ($V_{gs}$) around the conductance fluctuation at a temperature lower than 50 K. The minimum time resolution for the measurement was 0.01 s. Low temperature measurement is helpful for the observation of clear RTSs because RTSs are governed by a thermally activated process. We assume that the defect is the acceptor type, because typical $MoS_2$ crystal shows *n*-type and the defect which can communicate with carriers are usually close to conduction band, which has large probability to be acceptor [30]. Thus, simple two-level RTSs usually correspond to the capture and emission process for a single carrier with the shallow defect just below conduction band. The high current level corresponds to state in which a carrier is emitted, whereas a low current level corresponds to a state in which a carrier is captured, as shown in figure 2b. To confirm this scenario, statistical analysis of the RTSs with sufficiently long measurement time are studied, in which more than 300 transitions were included. The number of high current levels with the same capture time were counted and plotted as a function of capture time in figure 2c. Here, based on the probability distribution of an RTS [19], the probability $p(t)$ that the carrier will remain at the high level during a capture time of $t$ and then make one transition between times $t$ and $t$+d$t$ can be expressed by the following distribution:

$$p(t) = \frac{1}{t}\exp\left(-\frac{t}{\tau}\right),$$

where τ is the mean capture time for the high level. According to the $p(t)$–$t$ plot, the experimental data were reasonably fitted by this equation, which indicates that simple RTSs correspond to a single electron capture-emission process.

Since defects are randomly distributed in the channel and usually have a wide energy range, multilevel RTSs were also occasionally observed in one $I_{ds}$-Time measurement with constant $V_{gs}$ and temperature. Multilevel RTSs are discussed in this part. Most observed multilevel RTSs can be considered as the superposition of simple RTSs. For the simple RTSs, the amplitude is usually determined by how serious the defect affect conductive path. The short distance between the defect and conductive path corresponds to the large amplitude. Time constant of RTSs is usually determined by the barrier height between defect level and conduction band of $MoS_2$. The large barrier corresponds to slow RTSs. Therefore, RTSs in figure 3a can be understood as the superposition of the small RTSs and large RTSs, and RTSs in figure 3b can be regarded as the superposition of quick RTSs and slow RTSs. These RTSs are regarded as independent multilevel RTSs because they can be understood as the superposition of two or more simple, independent RTSs. In contrast, the RTSs in figure 3c are unlikely to be reproduced by merely superimposing two independent RTSs. The observation of quick RTSs at only the high current level suggests that two simple RTSs are correlated. These multilevel RTSs are regarded as correlated multilevel RTSs. Especially, correlated multilevel RTSs are quite rare and most of RTSs observed are quite sensitive to the gate voltage. Therefore, the gate voltage was severely adjusted to observe these RTSs.

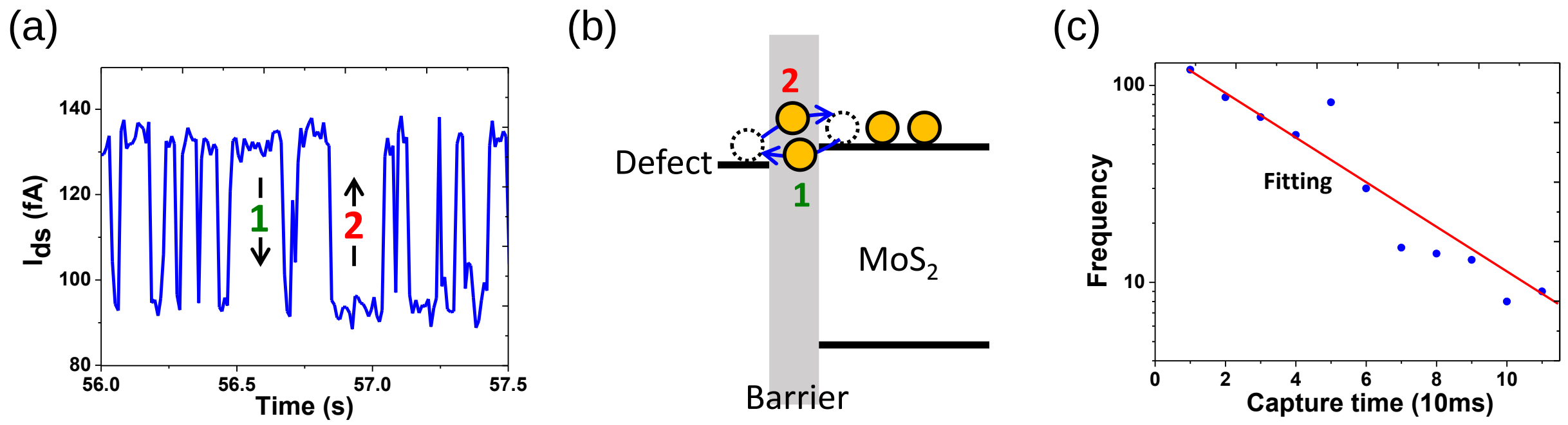

**Figure 2.** (a) $I_{ds}$ as a function of time for a bilayer $MoS_2$ FET measured at 20 K at $V_{gs}$= -2.1 V and $V_{ds}$ = 0.1 V. Simple two-level RTSs are observed. (b) Schematic drawing to explain the carrier capture and emission process by the defect. “1” and “2” indicate carrier trapped and emitted situations, respectively. The barrier height and width and the energy level difference between the defect and $MoS_2$ conduction band determines the lifetime of RTSs. The origin of the barrier can be a vacuum level or dielectric layer, which is determined by the positions of defects. Once the electron is trapped by the defect, carrier number reduction and scattering from the charged defect will reduce the current, facilitating a transition to a low current level. (c) The number of high current levels with the same capture time as a function of capture time.

The possible defect–defect interaction is considered to explain correlated multilevel RTSs. The correlated one is more frequently observed in the subthreshold region. This implies that the defect–defect interaction may result from the Coulombic interaction because this weak interaction might be screened in a high-carrier density region [31]. To understand the origin of correlated RTSs, two defects are considered. Assume that defect-A and defect-B correspond to quick and slow RTSs, respectively. The charged state of defect-B should affect that of defect-A, as shown in figure 4 [15,16]. It should be emphasized that both defects A and B are spatially and energetically within reach of each other. The atomic structure of $MoS_2$ has been studied by scanning transmission microscopy, which indicates that distance between defects cluster can be around 10nm or even less [26]. This result supports the relatively close distance between the defects. In the case of figure 4a, the carrier capture and emission between defect-A and the channel often occur if no carrier is captured in defect-B, which corresponds to a high level of RTSs. In contrast, in the case of figure 4b, when defect-B is charged, the energy level of defect-A is shifted upward, possibly because of the Coulombic interaction. This energy shift prevents carrier communication between defect-A and the channel. No RTSs are observed through defect-A. It should be noted that two types of defects A and B should be closely located to each other to facilitate the Coulombic interaction between two defects. The observation of correlated multilevel RTSs in $MoS_2$ FETs suggests that the defects affect carrier transport very seriously and that the screening in the atomically thin $MoS_2$ is weak compared with three-dimensional material.

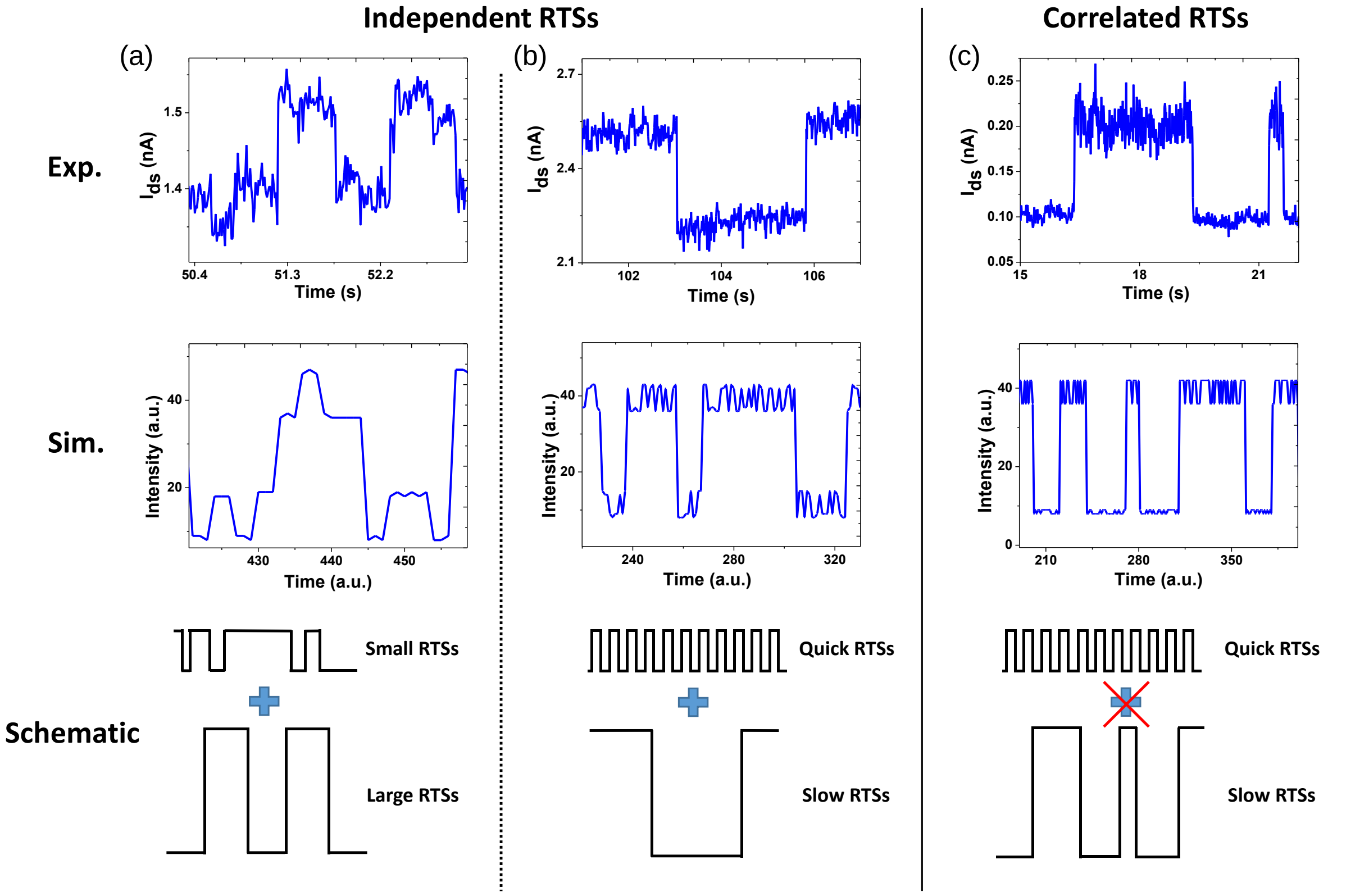

**Figure 3.** Classification of multilevel RTSs. The left two cases (a) and (b) are independent multilevel RTSs, whereas (c) correlated multilevel RTSs are on the right. (a) RTSs observed in monolayer $MoS_2$ FET at 50 K, $V_{gs}$ = -6.6 V and $V_{ds}$ = 0.1 V. This can be reproduced by the superposition of small RTSs and large RTSs. (b) RTSs observed in monolayer $MoS_2$ FET at 20 K, $V_{gs}$ = -6.3 V and $V_{ds}$ = 0.1 V. This can be reproduced by the superposition of quick RTSs and slow RTSs. (c) RTSs observed in monolayer $MoS_2$ FET at 50 K, $V_{gs}$ = 21.5 V and $V_{ds}$ = 0.1 V. This cannot be reproduced by the superposition of two simple RTSs. To reproduce these correlated multilevel RTSs, the defect–defect interaction due to Coulombic interaction should be considered.

A simulation using the random number for capture–emission process was conducted to reproduce various types of observed RTSs. Simple simple RTSs with specific interaction between two or more simple RTSs, which might be due to Coulombic interaction.

Thus far, we have discussed RTSs observed in pristine $MoS_2$ FETs and suggested that RTSs originate from the defect capture and emission process. Next, let us observe the RTS behavior for a $MoS_2$ channel intentionally irradiated by an electron beam (EB). It is known that EB irradiation can create defects inside $MoS_2$ [32-35]. Only monolayer $MoS_2$ is studied in this part. The EB (50 kV, 1.6 C/cm$^2$) was irradiated at the rectangular part in monolayer $MoS_2$ in figure 5a. No obvious morphology change was observed after EB irradiation, which was measured by AFM. In contrast, for the PL measurement, a new peak at RTSs correspond to one type of random number process, whereas multilevel RTSs can be simulated with the superposition of two independent random number processes. The superposition with specific conditions can lead to correlated RTSs. The simulated RTSs of independent and correlated multilevel RTSs are shown in figure 3. (Details of the simulation are shown in supplementary figure S3). The simulation can reproduce experimental data very well, which indicates that the correlated multilevel RTSs come from superposition of approximately 1.7 eV indicates that the defect level is created [36]. At the same EB irradiation condition, the $MoS_2$ channel was irradiated with a narrow line (figure 5c). The $I_{ds}$–$V_{gs}$ characteristics for this device are shown in figure 5d. As we expect, the carrier mobility is dramatically degraded after the EB irradiation. Interestingly, conductance fluctuation is clearly observed for narrow line patterned EB-irradiated devices at low temperature, which indicates that introduced defects can more seriously affect the carrier transport than defects in pristine $MoS_2$. The characterization of devices with other shapes patterned EB-irradiation is also shown in supplementary figure S4. The RTSs can be observed even in the on-state with a large $V_{gs}$ range, which is shown in supplementary figure S5. This intentionally created shallow defects have similar RTSs behavior (time constant and amplitude) with those observed in pristine $MoS_2$ devices, which also indicates that the energy level of defects in pristine $MoS_2$ might be identical to ones created by EB irradiation. From the new PL peak in figure 5b, the defect level may be located at 0.15 eV below conduction band. The $MoS_2$ channel with EB irradiation usually has a tendency to show RTSs in a large $V_{gs}$ range. This can be simply explained as a large number of defects being introduced in the $MoS_2$ channel.

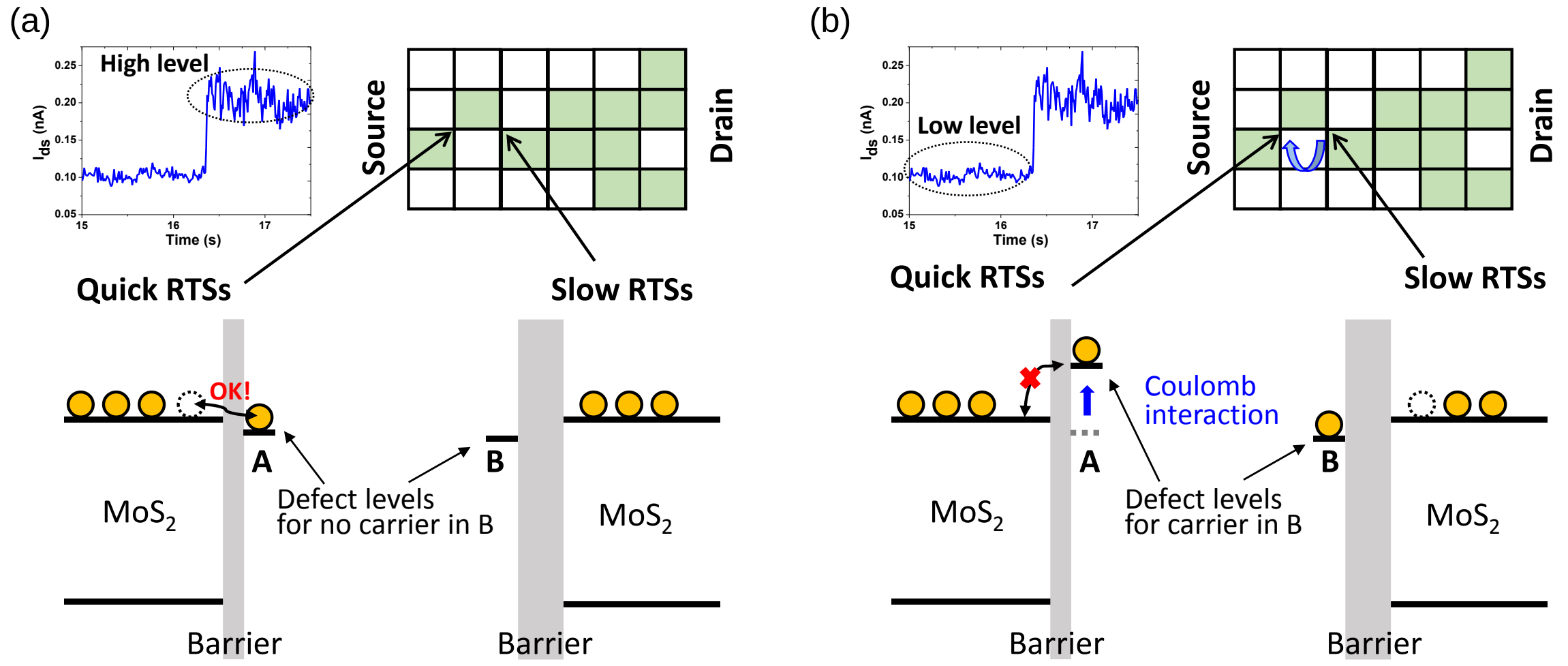

**Figure 4.** Schematic description of the model to understand the defect–defect interaction. (a) The case for high level with quick RTSs. The carrier capture and emission between defect-A and the channel is possible if no carrier is captured in defect-B. (b) The case for low level without quick RTSs. When defect-B is charged, the energy level of defect-A is shifted upward, possibly because of Coulombic interaction. This energy shift prevents carrier communication between defect-A and the channel, resulting in no quick RTSs in the low level.

$I_{ds}$ provides information on the whole channel region, not merely the local position. The resistance measurement at the local position by multi-probes allow the local positions of RTSs to be determined. Position detection is important for noise analysis because there is still controversy about the exact local position of $1/f$ noise in $MoS_2$ FET [10]. Figure 1a shows a device image of a measured thin $MoS_2$ FET. The minimal time resolution for the multi-probe measurement is set to 0.2 s because measured RTSs have a relatively long time constant. The synchronization for each source–measurement unit is confirmed at this time resolution. figure 6a, b shows a schematic drawing of a device with multi-probes and the resistance as a function of time for the whole channel region, $R_{ch}$, and local positions, $R_{S2}$, $R_{24}$ and $R_{4D}$, which are measured simultaneously. The resistance for all cases show RTS behavior. RTSs for each local resistance have the same time point. In other words, high and low levels of RTSs occur at the same time for each local resistance. Among them, the resistance between probes 2 and 4 shows the largest amplitude, which indicates that the positions of RTSs are located close to electrodes 2 and 4, as schematically shown in figure 6a. Interestingly, the effect of the defect is definitely far-reaching because RTS behavior is also observed in $R_{ch}$ and $R_{S2}$. Note that this device length is several micrometers, but the screening length in conventional three-dimensional systems is usually only nanometers [20,37]. This also indicates that the screening in $MoS_2$ is less effective than that in conventional three-dimensional systems. Another example is also shown in supplementary figure S6. The positions of RTSs are distributed in the whole channel.

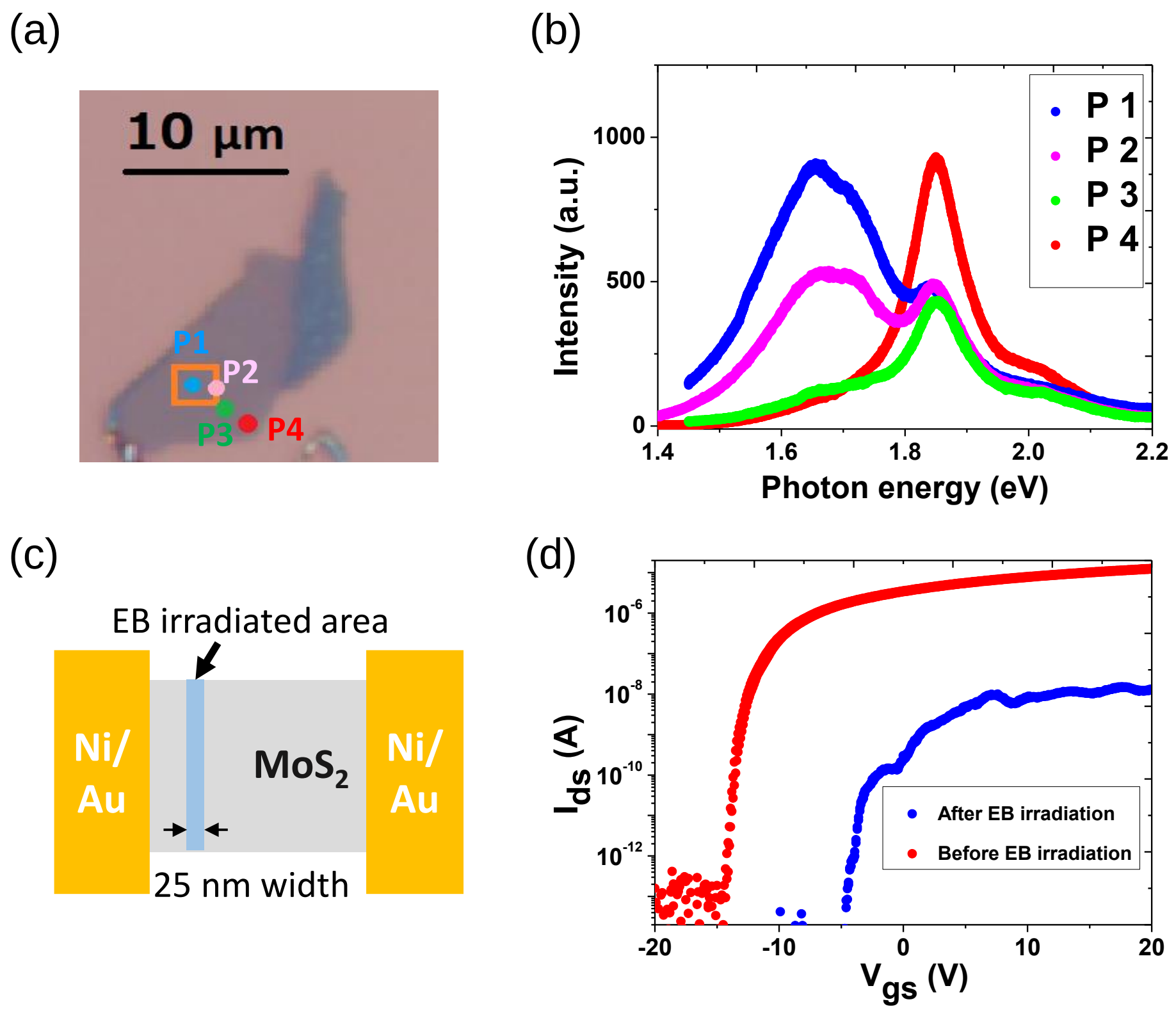

**Figure 5.** (a) Optical image of monolayer $MoS_2$ with EB irradiation. The area in the orange rectangle indicates the EB irradiated area. (b) PL data of different positions in monolayer $MoS_2$ with EB irradiation. The spots with different colors in (a) correspond to different PL curves. The peak at 1.85 eV corresponds to the direct bandgap of the monolayer, whereas the peak at 1.7 eV corresponds to defects introduced by EB irradiation. (c) Schematic drawing to show EB irradiated area for monolayer $MoS_2$ device. The line width for EB irradiation is 25 nm. (d) $I_{ds}$–$V_{gs}$ characteristics before and after EB irradiation for a monolayer $MoS_2$ FET at 50 K, $V_{ds}$ = 0.1 V. Carrier mobility is degraded after EB irradiation, and current fluctuation is observed in both the on-state and subthreshold region.

RTSs have been studied for more than half a century in Si-based devices, especially at the early stage. Regarding Si devices, the contact and the channel contributes to RTSs. RTSs, which result from contact, usually originate from the lattice imperfection of semiconductors close to contact. These defects including dislocation severely affect the carrier transport through the contact. With the development of process technology, including improvements in crystal quality and the implantation technology, these noises are rarely observed in current Si devices [8,38]. The channel RTSs come from defects not only in the semiconductor itself but also in the oxide layer and can still be occasionally observed in electronic devices. Regarding $MoS_2$, the present multi-probe experiments definitely support that the $MoS_2$ channel is one of the positions for RTSs. Moreover, in terms of the interface between the $MoS_2$ channel and the dielectric layer, the large interface states are reported, which also indicates the potential existence of channel RTSs [26,39]. Further, RTS study is very informative and promising to characterize the quality of a channel itself and other interfaces.

## 3. Conclusion

In conclusion, this paper presented a detailed analysis of RTSs in a thin layer $MoS_2$ transistor. RTSs were confirmed to result from the carrier capture and emission process. Independent multilevel RTSs can be understood as the superposition of simple ones. Correlated multilevel RTSs are understandable from the defect–defect interaction. The detection of RTSs in few-layer $MoS_2$ strongly suggests that the defects affect carrier transport very seriously and that screening in the atomically confined $MoS_2$ might be less effective than that in a conventional three-dimensional system. EB irradiation can introduce defects in $MoS_2$, which can also be characterized by RTSs. The local position for the occurrence of RTSs is studied by multi-probe measurement. Although they are distributed in the whole channel, the local positions of dominant RTSs can be detected. RTS study is very informative and promising to characterize defects inside $MoS_2$.

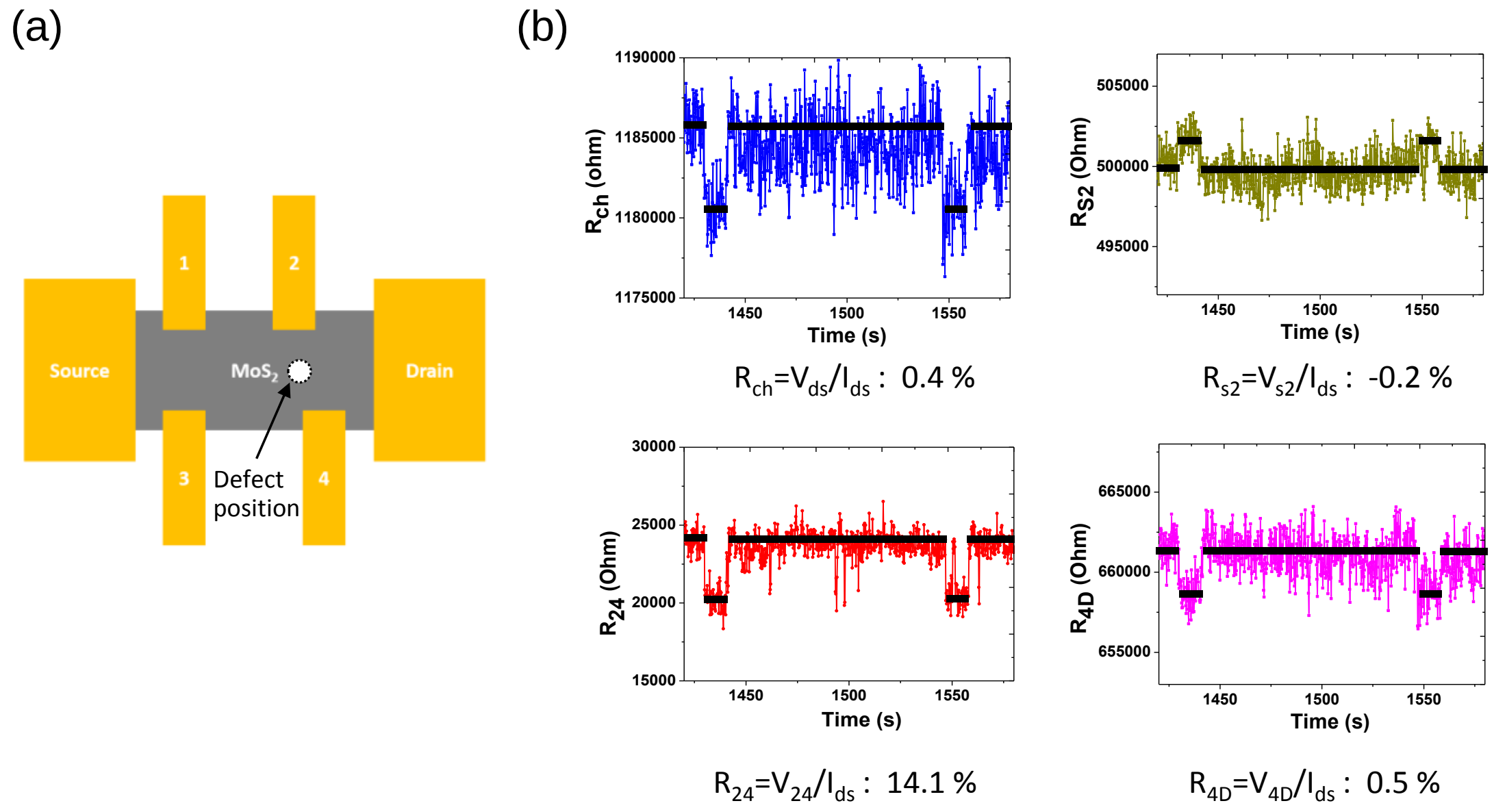

**Figure 6.** (a) Schematic to show positions of RTS origins. (b) Local resistance–time figure for each part, which were measured simultaneously. Temporal characteristics are measured at thin $MoS_2$ FET, $V_{gs}$ = -10 V, $V_{ds}$ = 4 V, and T = 10 K. Ti/Au was deposited for the measured multi-probe device. RTSs can be observed in all local parts. However, $R_{24}$ shows the largest amplitude, which indicates that RTS origins are around electrodes 2 and 4.

**Acknowledgements**
This work was partly supported by JST-CREST, JSPS Core-to-Core Program, A. Advanced Research Networks, and JSPS KAKENHI Grant Numbers JP25107004, JP16H04343, & JP16K14446.

**Supplementary data**

# Experimental detection of active defects in few layers $MoS_2$ through random telegraphic signals analysis observed in its FET characteristics

**Nan Fang, Kosuke Nagashio and Akira Toriumi**

Department of Materials Engineering, The University of Tokyo, Tokyo 113-8656, Japan

E-mail: nan@adam.t.u-tokyo.ac.jp

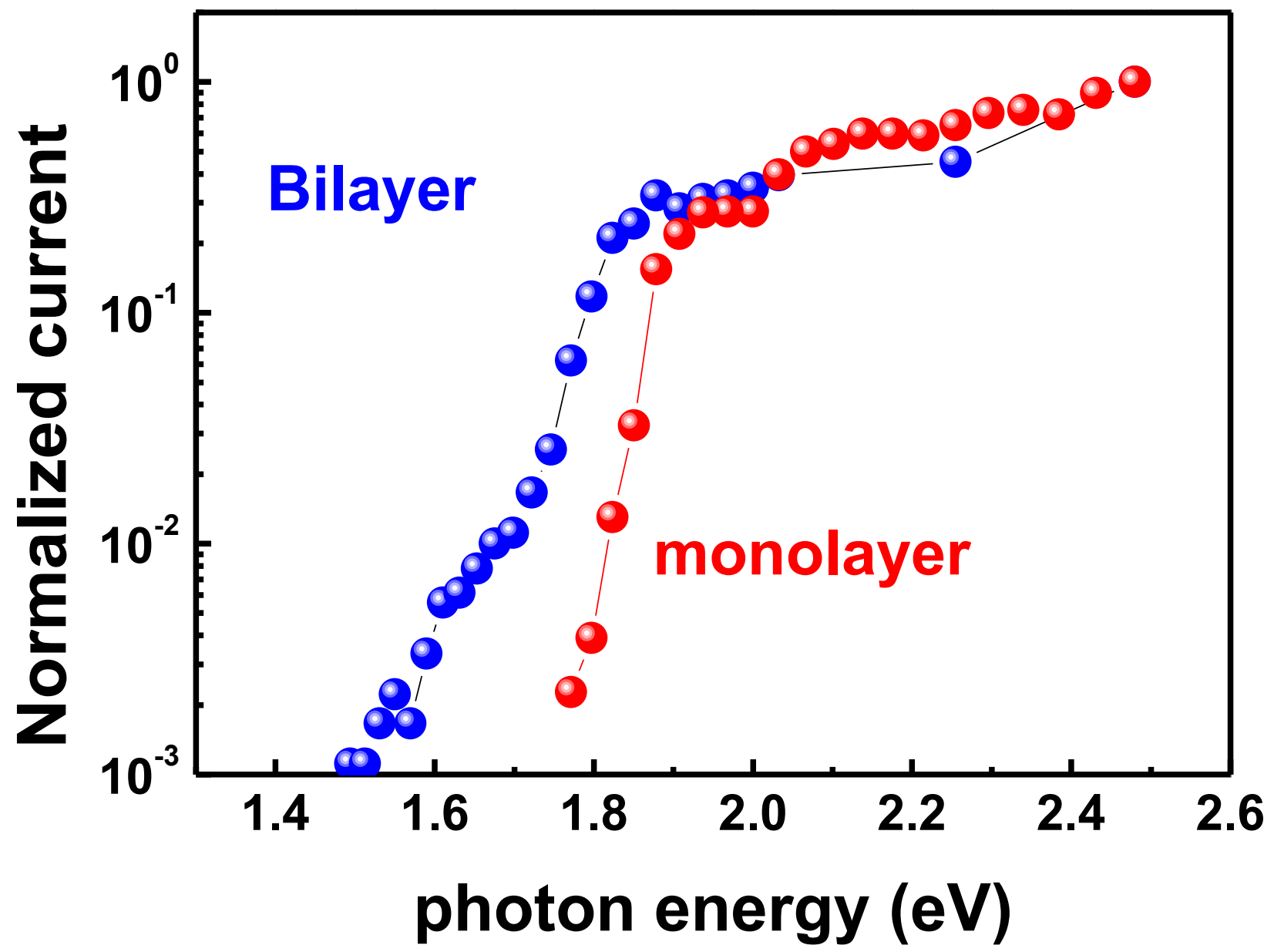

**Figure S1.** Photo current as a function of photon energy for monolayer (red curve) and bilayer (blue curve). Monolayer $MoS_2$ shows sharp photo current edge around 1.8eV, which corresponds to direct band-gap. While a shoulder around 1.6eV is observed in bilayer $MoS_2$, which corresponds to indirect bandgap.

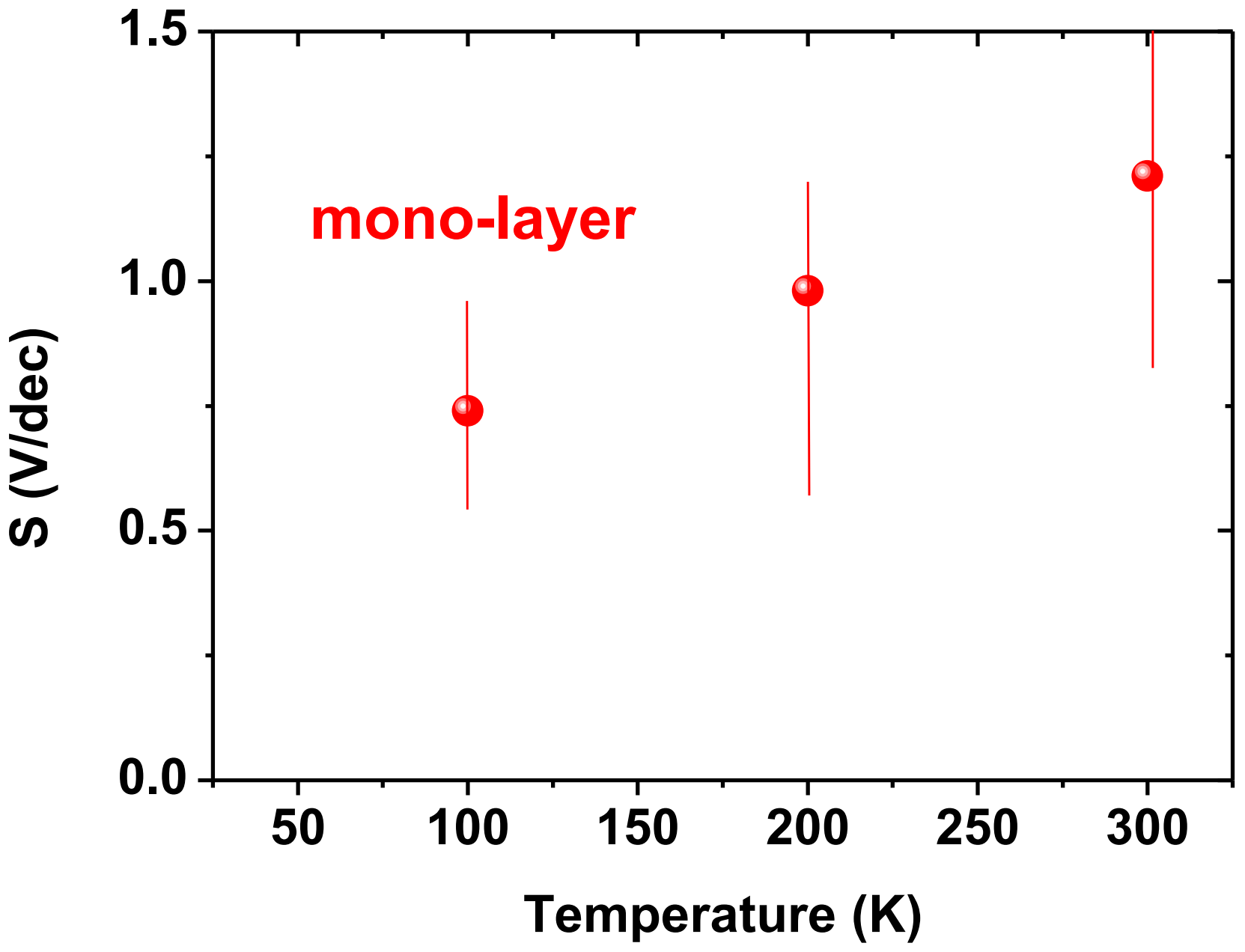

**Figure S2.** Dependence of subthreshold swing (S-factor) on measurement temperature. Almost linear temperature dependence is observed for monolayer $MoS_2$ FETs, which means the conventional method to analyze the subthreshold characteristics can be applied for $MoS_2$ FETs as well.

We extracted the interface states density from the temperature dependence by assuming that the contribution of depletion layer to the gate capacitance was negligible due to the atomically thin $MoS_2$ channel thickness. The interface states density, $N_x$, was estimated from the temperature dependence of the S-factor as follows.

$$\mathrm{Nx} = \int_{Ev}^{Ec} \frac{C_{ox}}{q} \left(S \frac{q}{k_B} \frac{1}{ln10} - 1\right) dE$$

$N_x$ was roughly estimated with $4\times10^{12}$ $cm^{-2}$ for more than 10 measured monolayer samples.

a

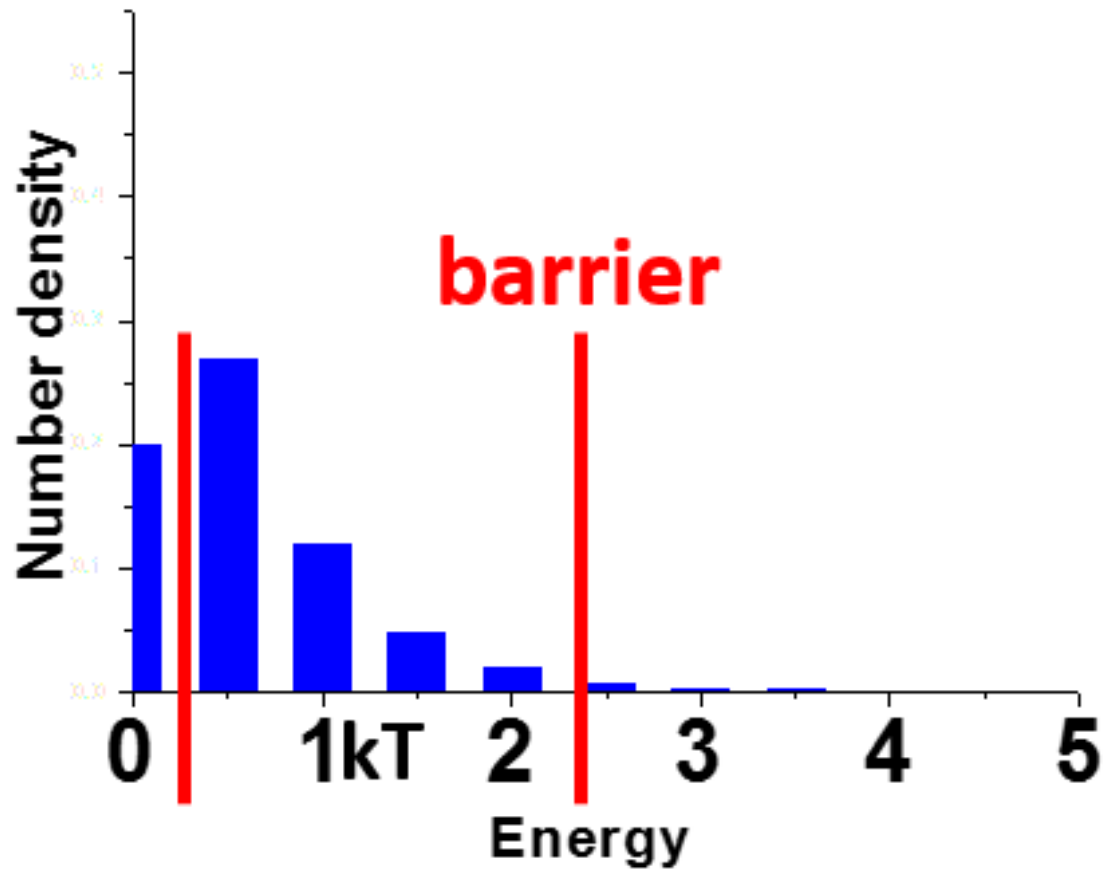

b

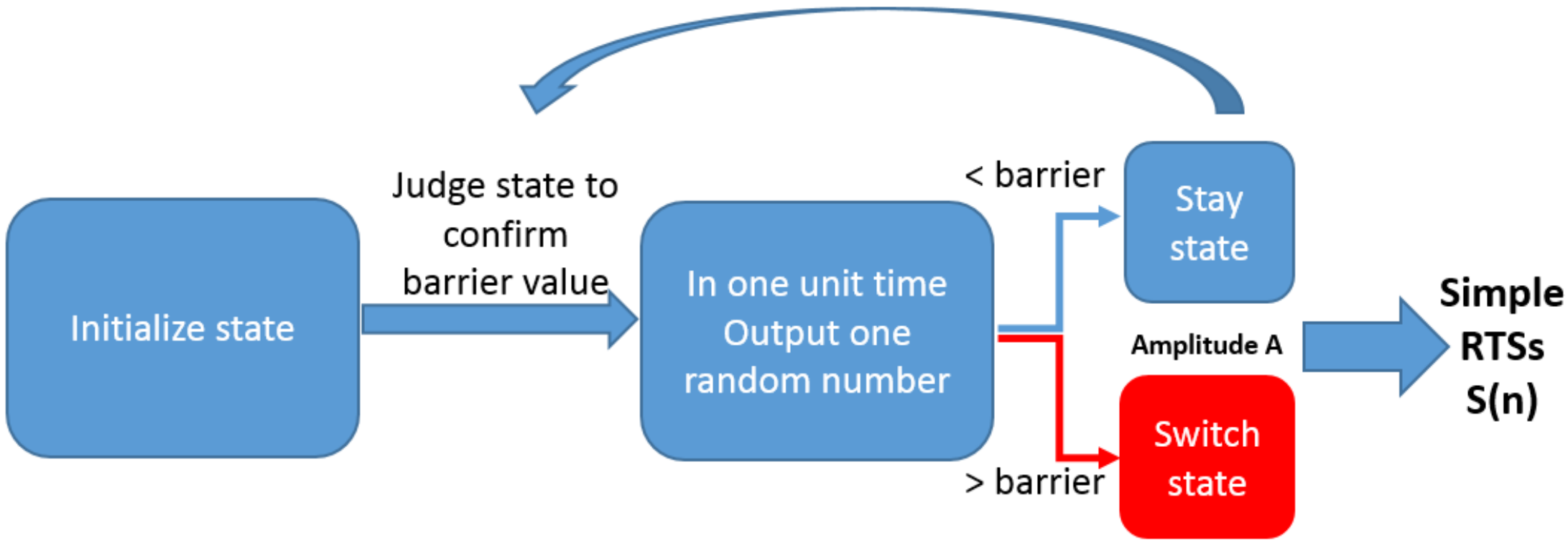

c

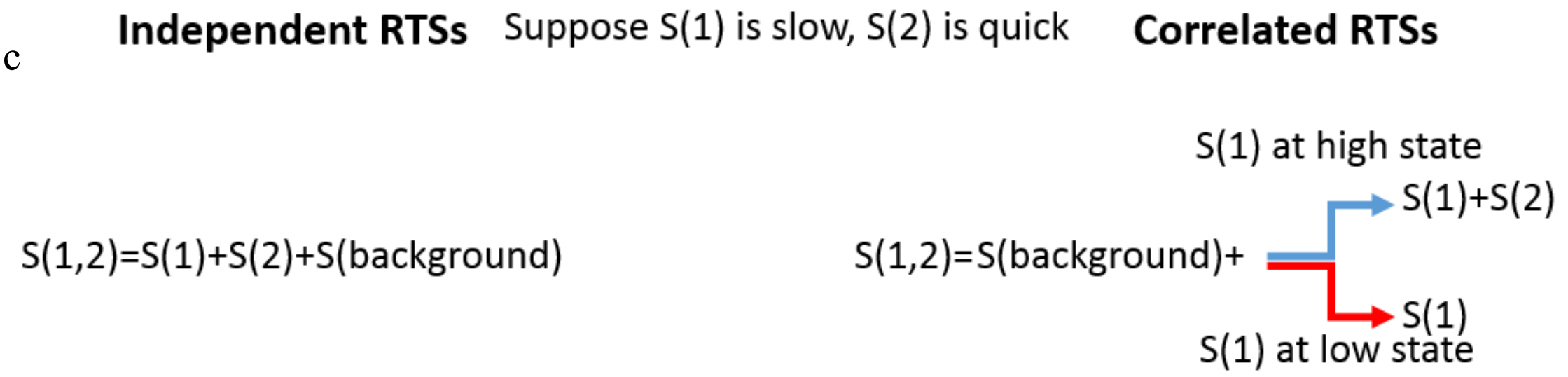

**Figure S3.** Simulation details of RTSs. (a) Exponential distribution of random number to simulate thermal process of electrons. Barriers are determined by capture and emission time, relatively. The long time indicates that high barrier, which makes the electron hard to overcome barrier. (b) Flow chart of simple RTSs simulation. S(n) is defined as simple RTSs with number n. (c) Conditions for superposition of simple RTSs to simulate multilevel RTSs including independent and correlated RTSs. Different types of multilevel RTSs can be simulated by changing the superposition conditions.

**Simulation of complex RTSs**

Firstly, based on amplitude A,$\tau_e$ and $\tau_c$ (suppose $\tau_c > \tau_e$ ), barriers for emission and capture process are estimated, relatively. The distribution of random number density (electron number) obeys exponential distribution to approximate the characteristics of the thermal equilibrium fluctuation on actual RTSs, although this is not an essential point in this simulation. The condition of determining each barrier is shown below.

- Amount of random numbers surmount 0/ total random numbers=1/$\tau_e$
- Amount of random numbers surmount energy barrier/ total random numbers= 1/ $\tau_c$

Then, based on flow chart in Figure S3b, single two-level RTSs can be estimated based on random number with the help of determined barrier. In one unit time if the value of extracted random number which obeys distribution in Figure S3a is larger than the barrier, the state of current will switch. If the value is smaller than the barrier, the state of current will stay. By repeating this process, simple two-level RTSs can be simulated.

Finally, different multilevel RTSs are simulated by different superposition of simple RTSs. Independent RTSs are just the superposition of simple RTSs without any other conditions while correlated RTSs are the superposition of simple RTSs with specific conditions. As shown in Figure S3c, S(2) is suppressed when S(1) is at low current level. By applying this specific condition, correlated RTSs are simulated, which fits experimental data very well. It indicates that the correlated RTSs come from superposition of simple RTSs with specific correlation.

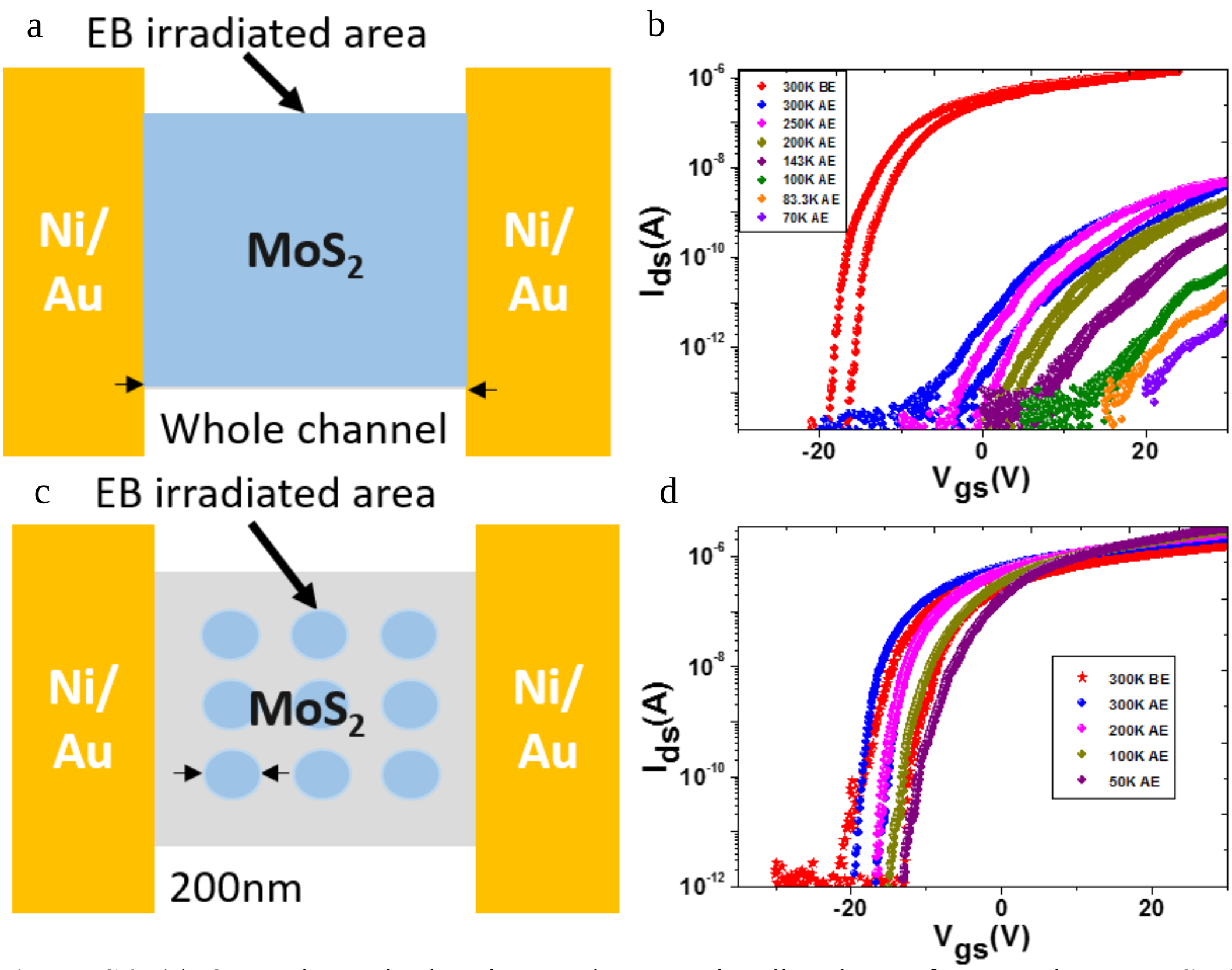

**Figure S4.** (a) One schematic drawing to show EB irradiated area for monolayer $MoS_2$ device. EB irradiation is conducted for the whole channel. (b) $I_{ds}$–$V_{gs}$ characteristics before (BE) and after EB irradiation (AE) for a monolayer $MoS_2$ FET (a) at different temperature, $V_{ds}$ = 0.1 V. Although carrier mobility is degraded after EB irradiation, neither large current fluctuation nor RTSs is observed. Moreover, $I_{ds}$ became too small to be measured below 50K. The absence of RTSs can also be understood by considering percolation type conduction. RTSs can only be observed when one single defect can seriously affect conductive path while one single defect cannot seriously affect conduction in case of the whole channel irradiation. (c) Another schematic drawing to show EB irradiated area for monolayer $MoS_2$ device. EB irradiation is conducted for the circle area with the diameter of 200 nm. (d) $I_{ds}$–$V_{gs}$ characteristics before and after EB irradiation for a monolayer $MoS_2$ FET (c) at different temperature, $V_{ds}$ = 0.1 V. Neither obvious carrier mobility degradation nor current fluctuation is observed after EB irradiation. It is because there is always conductive path between source and drain.

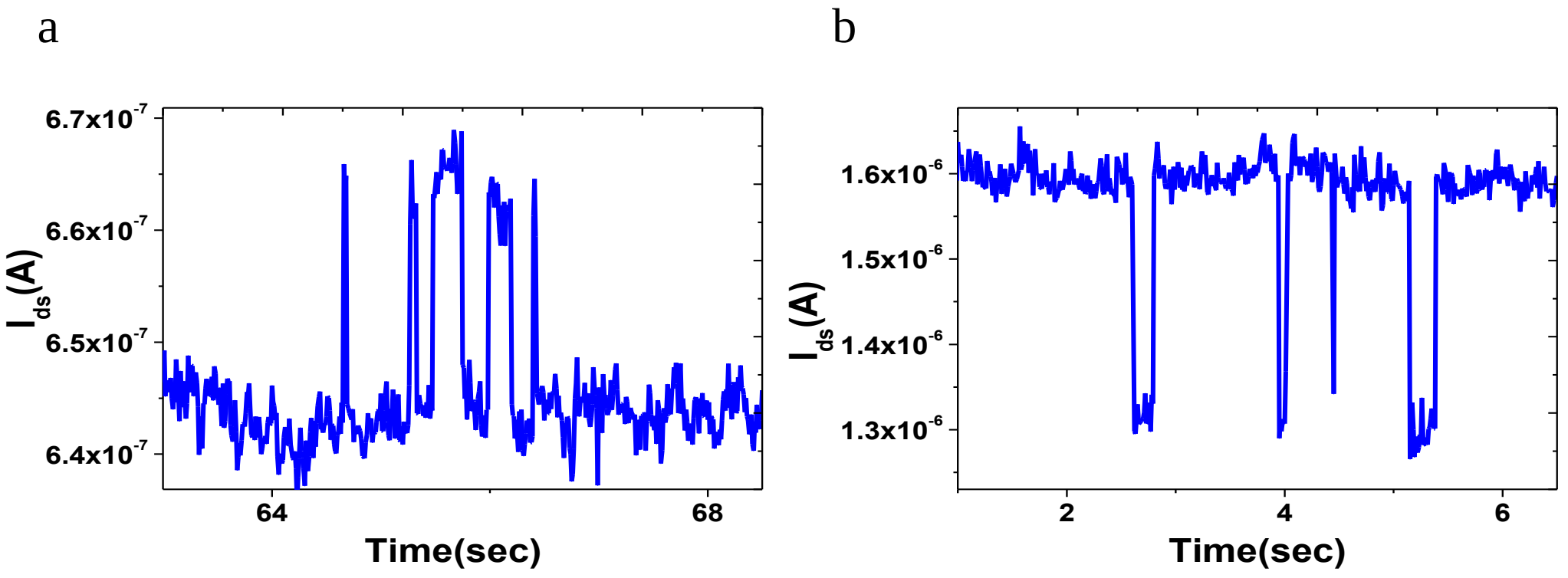

**Figure S5.** RTSs observed at on-state in EB-irradiated monolayer $MoS_2$ FET at 50 K $V_{ds} = 1$ V, (a) $V_{gs} = 5$ V (b) $V_{gs} = 10$ V.

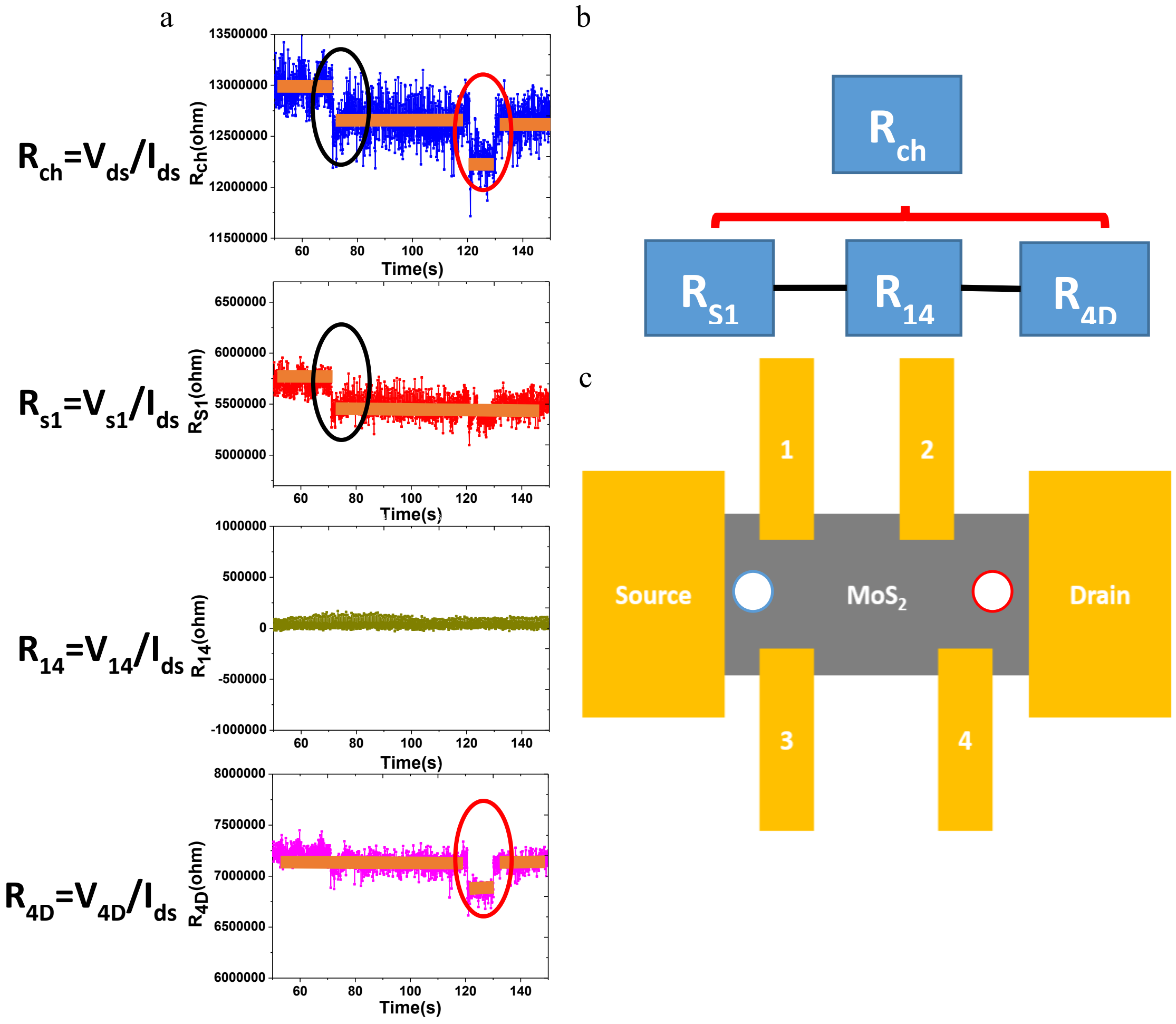

**Figure S6.** Another observed RTSs based on multi-probe measurement. Temporal characteristics are measured at thin $MoS_2$ FET, $V_{gs}$ = -3.1 V, $V_{ds}$ = 0.1 V, T = 10 K. Ti/Au was deposited for measured multi-probe device. (a) Local resistance-time figure for each part. RTSs can be observed in the local resistances. The black circled RTSs come from $R_{s1}$ while red circled RTSs come from $R_{4D}$. (b) Channel resistance consists of series of local resistances. (c) Schematic to show positions of RTSs origins. Both contact and channel from source to electrode 1 can contribute to the observed RTSs of $R_{S1}$ since $R_{S1}$ is a series of contact resistance and channel resistance from source to electrode 1.